\documentclass[aps,prd,showpacs,twocolumn,superscriptaddress,nofootinbib,preprintnumbers]{revtex4-1} 

\makeatletter
\def\p@subsection{}
\makeatother

\usepackage{color,graphicx,amsmath,amssymb}
\usepackage[colorlinks=true,citecolor=blue,linkcolor=blue,urlcolor=blue, backref=false,pdfborder={0 0 0}]{hyperref}

\newcommand{\be}{\begin{equation}}
\newcommand{\ee}{\end{equation}}
\newcommand{\beqa}{\begin{eqnarray}}
\newcommand{\eeqa}{\end{eqnarray}}

\newcommand\m{\mu}

\newcommand\x{\textbf x}

\newcommand\g{\gamma}

\newcommand\n{\nu}
\renewcommand\r{\rho}
\newcommand\s{\sigma}

\renewcommand\l{\lambda}

\def\e{{\rm e}}
\def\d{\partial}
\newcommand{\bseq}{\begin{subequations}}
\newcommand{\eseq}{\end{subequations}}

\newcommand{\di}{\mathrm d}

\def\gsim{\raise0.3ex\hbox{$\;>$\kern-0.75em\raise-1.1ex\hbox{$\sim\;$}}}
\def\lsim{\raise0.3ex\hbox{$\;<$\kern-0.75em\raise-1.1ex\hbox{$\sim\;$}}}

\def\beqn#1{\begin{equation}\label{#1}}
\def\eeqn{\end{equation}}

\def\beqa#1{\begin{eqnarray}\label{#1}}
\def\eeqa{\end{eqnarray}}

\def\Z2{$\mathcal{Z_2}$}

\newcommand {\ignore}[1]{}

\begin{document}

\preprint{INR-TH-2019-017}
\preprint{KCL-PH-TH/2019-74}

\title{No chiral light bending by clumps of axion-like particles}



 \author{Diego Blas}\affiliation{Theoretical Particle Physics and Cosmology Group, Department of Physics,\\
   King's College London, Strand, London WC2R 2LS, UK}
\author{Andrea Caputo}\affiliation{Instituto de Fisica Corpuscular, Universidad de Valencia and CSIC, Edificio Institutos Investigacion, Catedratico Jose Beltran 2, Paterna, 46980 Spain}
\author{Mikhail M. Ivanov}\email{mi1271@nyu.edu}\affiliation{Center for Cosmology and Particle Physics, Department of Physics, New York University, New York, NY 10003, USA}
\affiliation{Institute for Nuclear Research of the
Russian Academy of Sciences, \\ 
60th October Anniversary Prospect, 7a, 117312
Moscow, Russia
}
\author{Laura Sberna}\affiliation{Perimeter Institute, 31 Caroline St N, Ontario, Canada}

\begin{abstract} 
We study the propagation of light 
in the presence of a parity-violating coupling between photons and axion-like particles (ALPs).
Naively, this interaction could lead to a split of light rays into two separate beams of different polarization chirality
and with different refraction angles. 
However, by using the eikonal method 
we explicitly show that this is not the case and that ALP clumps
do not produce any spatial birefringence. 
This happens due to non-trivial variations 
of the photon's frequency and wavevector, which absorb time-derivatives and gradients of the ALP field.
We argue that these variations represent a new way to probe the ALP-photon couping with 
precision frequency measurements.
\end{abstract}

\maketitle


\section{Introduction}

Astrophysical data accumulated over the past decades provide an overwhelming evidence 
for the existence of a cold matter component of non-baryonic origin. The leading explanation 
 is to complete the Standard model of particle physics (SM) with new particles, referred to as dark matter (DM). 
 
Regardless of astrophysical data, the SM has its own puzzles, which make DM candidates addressing them particularly appealing.  
One such candidate is the quantum -- chromodynamics (QCD) axion, 
which  appears in the Peccei-Quinn solution to the strong CP problem  \mbox{\cite{Peccei:1977hh, Peccei:1977ur, Weinberg:1977ma, Wilczek:1977pj, Preskill:1982cy, Dine:1981rt}}. 
More generically, many
extensions of the SM  predict light particles with similar properties to the QCD axion, but unrelated to the strong CP problem. These are known as axion-like particles (ALPs), and
appear  in particular in low energy-effective theories of string theory~\cite{Witten:1984dg, Arvanitaki:2009fg}. In the following we will refer to both cases as ALPs. 

Another motivation for  ALPs is that they may resolve the difficulties of  heavier DM candidates 
 to explain observations on galactic scales, see e.g. \cite{DelPopolo:2016emo,Hui:2016ltb}. This topic has recently attracted significant attention in the context of
the so-called fuzzy dark matter, a hypothetic ALP with extremely light 
mass $m\sim 10^{-22}$ eV \cite{Hu:2000ke,Hui:2016ltb}.
Different observations disfavor these extremely small masses, while larger masses are still viable  \cite{Bar:2018acw,Lancaster:2019mde,Marsh:2018zyw,Nori:2018pka}.

A peculiarity of the ALP field (denoted by $a$) is its coupling to photons, 
\be 
\label{eq:Lag}
\mathcal{L}_{\text{int}}=\frac{g_{a\g\g}}{4}aF_{\mu\nu}\tilde{F}^{\mu\nu}\,,
\ee
where $\tilde{F}_{\mu\nu}=\frac{1}{2}\epsilon_{\m\n\r\s}F^{\r\s}$, $F_{\mu\nu}=\d_{\mu} A_{\nu}-\d_\nu A_\mu$ and $A_\mu$ is the photon field. Many observational effects of this interaction have been scrutinized in various 
astronomical and experimental settings \cite{Hagmann:1990tj, Hagmann:1998cb, Espriu:2011vj,Andrianov:2013bxa,Espriu:2014lma,Du:2018uak, Zhong:2018rsr, Armengaud:2014gea, Raffelt:2012sp,Kahn:2016aff, Graham:2011qk, Budker:2013hfa, Barbieri:1985cp, Flambaum:2018ssk, Stadnik:2017hpa, Caputo:2018ljp, Sigl:2017sew, Hook:2018iia, Safdi:2018oeu, Rosa:2017ury, Caputo:2018vmy, Caputo:2019tms, Harari:1992ea, Liu:2019brz, Sigl:2018fba, Ivanov:2018byi, Fedderke:2019ajk,Liu:2018icu,DeRocco:2018jwe,Obata:2018vvr}. 

Naively, the coupling \eqref{eq:Lag}
suggests that an ALP background might act as a usual 
optically active chiral medium. 
Recall that when a linearly polarized light beam 
enters such a medium, it splits into two rays of different chirality. 
The difference between refractive indices makes these waves 
acquire different phases as they propagate though the chiral fluid.  
Importantly, the two rays are \textit{spatially separated} and propagate independently from each other.
If the waves are recombined at the observation site, 
the resulting light should exhibit a rotation of the polarization plane as compared to the initial state.
This effect is known as ``natural optical rotation''~\cite{landau2013electrodynamics}.

The presence of the polarization plane rotation 
induced the ALP-photon coupling was originally pointed out by 
Harari and Sikivie in Ref.~\cite{Harari:1992ea}.
Remarkably, for the coupling \eqref{eq:Lag} the 
optical rotation 
depends only on the initial and final configurations
of the ALP field \cite{Harari:1992ea,Fedderke:2019ajk}. 
Pushing the analogy 
between the ALP field and a chiral medium, one might expect 
that inhomogeneities of the ALP field can generate a spatial split (``spatial birefringence'') 
between two photon's polarizations.

Since the paper by Harari and Sikivie, anomalous optical properties of 
the ALP fluid were studied in many works, but only few of them 
discussed the spatial separation of the photons \cite{Espriu:2011vj,Andrianov:2013bxa,Espriu:2014lma,Plascencia:2017kca}. 
In the present paper we fill this gap 
and scrutinize the effect of light deflections by ALP clumps. To this end we derive the 
light tracing equation with the eikonal method, which is valid
for generic (mildly) time-dependent and refractive media.

\section{Eikonal calculation}

Consider the following Lagrangian density for an ALP of mass $m$ interacting with photons:
\be
\mathcal{L}=-\frac{1}{4}F_{\mu\nu}^2 +\frac{1}{2}(\d_\mu a \d^\mu a - m^2 a^2) +\frac{a g_{a\g\g}}{4} F_{\mu\nu}\tilde{F}^{\mu\nu}\,.
\ee
The equations of motion are\footnote{Note that $\epsilon^{0123}=-\epsilon_{0123}=1$, and we use the following 
shorthands: $\Delta=\d_i\d^i$\,, $\Box=\d^\mu \d_\mu$.}
\be
\begin{split}
&\Box A_\nu -\d_\n\d_\m A_\m - g_{a\g\g}\epsilon^{\m\n\l\r}\d_{\m}(a \,\d_\l A_\r) = 0\,, \\
& - \Box a -m^2 a + \frac{g_{a\g\g}}{4} F_{\mu\nu}\tilde{F}^{\mu\nu} = 0\,.
\end{split}
\ee
Introducing the electric and magnetic components as
$F_{ij}=-\epsilon_{ijk}H_k\,,\quad F_{0i}=E_i$,
we find the following pair of modified Maxwell equations,
\begin{align}
	\label{eq:Maxwell}
& \d_0 E_i - \epsilon_{ijk}\d_j H_k +g_{a\g\g} \left[\d_0(a H_i)+\epsilon_{ijk} \d_j (a E_k)\right]=0\,,\nonumber \\
& \d_i E_i +g_{a\g\g}\d_i(a H_i)=0\,.
\end{align}
The Bianchi identities take the standard form,
\be
\begin{split}
 \d_i H_i=0\,,\quad \d_0 H_i+\epsilon_{ijk} \d_j E_k=0\,,
\end{split}
\ee
while the ALP equation of motion reads
\be 
\label{eq:axfull}
(-\d_0^2 + \Delta -m^2)a - g_{a\g\g} E_i H_i =0\,.
\ee
We study the propagation of electromagnetic fluctuations in a {\it fixed} ALP astrophysical background.
In particular, we ignore any back-reaction from the photons on the background.  
For simplicity, the results of this paper are derived assuming a flat Minkowski metric.

We will work in the ray optics (eikonal) approximation,  
\be 
\label{eq:adiab}
\left|{\d_0 a}/{a}\right|,\,\,\, \left|{\d_i a}/{a}\right| \ll \omega \,,
\ee
where $\omega$ is the photon frequency. 
Note that we do not assume any hierarchy between 
the temporal and spatial variations of the ALP field and keep all powers of $g_{a\g\g}$.
From \eqref{eq:adiab} we can neglect terms with two derivatives acting on $a$ and  find the following wave equations:
\be
\label{eq:Max}
\begin{split}
& (\d_0^2 - \Delta )E_i 
+g_{a\g\g}\left[\d_0 a\d_0H_i 
-\d_ja\d_j H_i\right]=0\,,\\
& (\d_0^2 - \Delta )H_i - g_{a\g\g}\left[\d_0a\d_0 E_i 
-\d_ja \d_j E_i\right]
=0\,.
\end{split}
\ee
To derive the photon path we follow the general eikonal formalism of Ref.~\cite{PhysRev.126.1899}.
First, we rewrite the wave equations
\eqref{eq:Max} in the following operator form:
\be
\hat{M}(t,x^i)*(\vec{E},\vec{H})^{\text{T}}= 0\,.
\ee
The zeroth order eikonal approximation amounts to using an ansatz 
$E_i,H_i\propto \e^{iS}$,
where the phase $S$ formally defines the photon's frequency
and wavevector along the ray orbit as 
\be
\omega(t,x^j(t)) \equiv -\d_0 S \,,\quad k_i(t,x^j(t))\equiv \d_i S \,.
\ee
The eigenvalues of the operator $\hat{M}$ are given by
\be
\begin{split}
D^{\pm}_{0}=\omega^2-k^2\pm g_{a\g\g}(\omega\d_0 a+k^i \d_i a)\,,
\end{split} 
\ee
where $\pm$ corresponds to right and left circular polarizations.
Using the standard equations for the photon's orbit,
\be
\begin{split}
& \frac{\di x^i}{\di t}=-\frac{{\d D_0}/{\d k^i}}{{\d D_0}/{\d \omega}}\,,\quad \frac{\di k^i}{\di t}=\frac{{\d D_0}/{\d x^i}}{{\d D_0}/{\d \omega}}\,,\\
& \frac{\di \omega}{\di t}=-\frac{{\d D_0}/{\d t}}{{\d D_0}/{\d \omega}}\,, \label{eq:domega}
\end{split}
\ee 
we find:
\bseq
\label{eq:eikall}
\begin{align}
\label{eikx}
& \frac{\di x^i}{\di t}=\frac{k^i \mp g_{a\g\g}\d_i a/2}{\omega \pm g_{a\g\g}\d_0a/2}\,,\\
\label{eikk}
& \frac{\di k^i}{\di t}=\pm g_{a\g\g}\frac{\omega \d_i\d_0 a + k^j \d_j \d_i a}{2\omega \pm g_{a\g\g}\d_0a}\,,\\
\label{eikomega}
& \frac{\di \omega}{\di t}=\mp g_{a\g\g}\frac{\omega \d_0^2 a + k^i \d_i \d_0 a}{2\omega \pm g_{a\g\g}\d_0a}\,.
\end{align}
\eseq
Taking a second total time derivative of \eqref{eikx}
and using Eq.~(\ref{eq:eikall}) we arrive at
\be
\label{eq:final}
\begin{split}
\frac{\di^2 x^i}{\di t^2}=0\,,
\end{split}
\ee
which formally holds true to all orders\footnote{Note that at order $\mathcal{O}(g^2_{a\g\g})$ corrections to geometric optics (which we did not consider) are formally not negligible.} in $g_{a\g\g}$. 
This implies that light is not bent by the ALP clumps.
We stress that the result \eqref{eq:final} holds true \textit{for any}
ALP configuration, even for unphysical time-independent profiles.

Importantly, Eq.~\eqref{eq:final} dictates that
\be 
\label{eq:ndef}
\left|\frac{\di x^i}{\di t}\right|^2 = \frac{|k^i\mp g_{a\g\g}\d_i a/2|^2}{(\omega \pm g_{a\g\g}\d_0a/2)^2}=1\,,
\ee
which means that the variations of the photon's frequency and 
wavevector identically cancel time derivatives and gradients of the ALP field. 
This implies that the photon's frequency and wavevector must vary between the observer
and the source sites. The total variation is obtained by
integrating
Eqs.~\eqref{eikk} and ~\eqref{eikomega} along the ray path 
(keeping terms of order $\mathcal{O}(g_{a\g\g})$),
\bseq
\label{eq:var}
\begin{align}
\label{vark}
& \Delta k^i=\pm \frac{g_{a\g\g}}{2} (\d_i a(t_{\text{d}},\x_{\text{d}})- \d_i a(t_{\text{e}},\x_{\text{e}}))\,,\\
\label{varom}
& \Delta \omega=\mp \frac{g_{a\g\g}}{2} ( \d_0 a(t_{\text{d}},\x_{\text{d}})- \d_0 a(t_{\text{e}},\x_{\text{e}}))\,,
\end{align}
\eseq
where the subscripts ``e'' and ``d'' denote the moments of emission 
and detection, respectively.
This implies that oscillations of the ALP field with period $2\pi/m$ induce an oscillating component in the photon's 
frequency, which varies with the same period (see also \cite{Carroll:1991zs}).
This effect is suppressed 
by $m/\omega$ compared to the polarization plane rotation (Harari-Sikivie effect) \cite{Harari:1992ea}. 
The typical frequency shifts for the hypothetical Galactic ALP can be crudely estimated as
\be
\label{eq:dw}
\frac{\Delta \omega}{\omega} \sim 10^{-16}\left(\frac{g_{a\g\g}}{10^{-10} \text{\small GeV}^{-1}}\right)
 \left(\frac{1\,\text{\small GHz}}{\omega}\right)\sqrt{\frac{\rho_{\text{DM}}}{0.3\,\text{\small GeV/cm}^3}} \,,
\ee
where $\rho_{\text{DM}}=m^2a^2/2$ is the local dark matter density. 
Such relative frequency shifts are similar to the accuracy of modern atomic clocks \cite{2014Metro..51..108G}. 
This suggests a possible new way to use atomic clocks to constrain 
ALPs, complementary to  \cite{Derevianko:2013oaa,Arvanitaki:2014faa,Stadnik:2015upa,Hees:2016gop,Wolf:2018xlz,Alonso:2018dxy}. 
An even stronger effect can be generated by a passage of a dense ALP minicluster 
through Earth (see Ref.~\cite{Pospelov:2012mt} for a similar idea in the context of domain walls).
We leave the exploration of these possibilities for future work.

\section{Conclusions}

We have shown that the ALP-photon coupling 
does not lead to light bending.
This result does not seem trivial to us, 
as it requires taking into account the 
variations of photon's frequency and wavevector, which are produced by 
fluctuations of the ALP field in space and time. 
The vanishing of the r.h.s. in Eq.~\eqref{eq:final} is a result of a delicate cancellation between these effects.
On a positive note, our analysis justifies 
the approach of previous works on the 
ALP-induced polarization plane rotation, which explicitly neglected 
all effects related to photon's refraction through the axion clumps.

The chiral deflections by ALP clumps were studied in 
Refs.~\cite{Plascencia:2017kca,Chway:2019prm}, 
though their results were derived using the assumption that the photon's frequency
is conserved. From Eq.~\eqref{eq:domega}, one sees that this assumption is not justified.

Finally, we have remarked that the absence of spatial birefringence implies that 
the photon's
frequency and wavevector must exhibit time-periodic components. 
We plan to explore their possible influence in very precise  frequency measurements in the future.
\vspace{1cm}

 \textit{Acknowledgments.}
We are grateful to 
Yacine Ali-Haimoud,
Masha Baryakhtar, 
Liang Dai,
Sergei Dubovsky,
Daniel G. Figueroa, 
Mehrdad Mirbabayi,
Alexandr Panin, 
Oriol Pujol\`as,
Javier Redondo,
George Raffelt, 
Benjamin Safdi, 
Andrei Shkerin, 
Inar Timiryasov,
Sergey Troitsky, 
Ken Van Tilburg,
Alfredo Urbano,
and Tomer Volansky for fruitful discussions. 
We thank Sergey Sibiryakov for valuable comments on the preliminary draft of this paper.
A.C. acknowledges support from national grants FPA2014-57816-P, FPA2017-85985-P and the European projects H2020-MSCAITN-2015//674896-ELUSIVES and H2020-MSCA-RISE2015.
The research of L.S. at Perimeter Institute is supported by the Government of Canada through Industry Canada and by the Province of Ontario through the Ministry of Research and Innovation. 
The work of M.I. on deriving the eikonal equations is supported by the Russian Science Foundation \mbox{(grant 18-12-00258)}.

\bibliography{biblio.bib}

\end{document}